\documentclass[conference,10pt,letterpaper]{IEEEtran}
\newcommand{\mao}[1]{\textcolor{orange}{Mao: #1}\PackageWarning{Mao:}{#1!}}
\renewcommand{\mao}[1]{}

\usepackage{graphicx} 
\usepackage[hidelinks,bookmarks=false]{hyperref}
\usepackage{booktabs}
\usepackage{xcolor}
\usepackage{amsmath, amssymb}
\usepackage[linesnumbered,ruled,vlined]{algorithm2e}
\usepackage{algpseudocode}
\usepackage{subcaption}  
\usepackage{comment}

\usepackage{graphicx} 

\title{Adaptive Swin Transformer Partitioning over AI-RAN Networks}
\author{
\IEEEauthorblockN{\small{Tam Thanh Nguyen, Yong Hao Pua, Tuan Van Ngo,
Mao V. Ngo, Jihong Park, Binbin Chen, and Tony Q.~S.~Quek}} 
\IEEEauthorblockA{{\small \{{nguyen\_thanhtam}, {yonghao\_pua}, {vantuan\_ngo}, {vanmao\_ngo}, {jihong\_park}, {binbin\_chen}, tonyquek\}@sutd.edu.sg}}} 
\date{March 2025}

\begin{document}
\maketitle
\begin{abstract}

This paper demonstrates the feasibility of transformer-based split inference for real-time video object detection over dynamic 5G AI-RAN networks. We extend throughput-aware adaptive splitting from CNNs to a Swin Transformer backbone and show that practical split execution is achievable for transformer-based vision models without retraining. To address the large intermediate activations inherent to transformers, we introduce an efficient, accuracy-preserving activation compression pipeline that substantially reduces uplink payload. The complete system—including adaptive split selection, transformer inference, and compression—is implemented and validated end-to-end on a real-time detection workload, with distributed UPF (dUPF) integration further reducing user-plane latency and improving runtime stability. Extensive measurements on an NVIDIA Aerial–based AI-RAN testbed jointly account for inference and 5G communication energy, quantifying the latency–energy–privacy trade-offs in realistic deployments.

\end{abstract}


\section{Introduction}


\maketitle

Mobile devices increasingly run compute-intensive AI inference for applications such as real-time video understanding, AR/VR, and intelligent monitoring. Executing full deep neural networks (DNNs) on UE often causes high latency and rapid battery drain, while offloading entire workloads to edge/cloud can introduce privacy exposure and network-dependent delays. Split computing (SC) offers a practical middle ground by partitioning the model: early layers (the \textit{head}) run on the UE and intermediate features are transmitted to an edge server that completes inference (the \textit{tail}).

However, most deployed SC systems rely on fixed partitioning, which becomes inefficient in time-varying 5G environments. In practice, throughput can fluctuate due to interference, mobility, scheduling contention, and other dynamics. Under such conditions, a fixed splitting point can lead to end-to-end (E2E) delay spikes, wasted UE energy (either from excessive local computation or repeated transmissions), and weakened privacy protection if the system is forced to transmit more “raw” intermediate features to meet latency constraints.

To address this, our work advances an adaptive AI-on-RAN framework that continuously senses the radio environment and adjusts the model split in real time. Building on our earlier work~\cite{TamGlobecom2025} of adaptive partitioning driven by AI-powered throughput estimation, we extend the framework to support Swin Transformer~\cite{liu2021swin}—a mainstream vision backbone whose hierarchical structure enables flexible partitioning options beyond traditional CNNs. Our system targets privacy-sensitive real-time video object detection and integrates with dUPF to further reduce E2E latency in edge AI pipelines.

A key requirement for optimal splitting is accurate knowledge of the maximum achievable uplink throughput for transmitting intermediate features. Yet, throughput estimation from conventional radio measurements can be unreliable, especially under challenging interference patterns. In our prior study~\cite{TamGlobecom2025}, we observed that standard numerical KPMs can fail to characterize throughput drops in certain regimes; \textit{augmenting with IQ-derived spectrogram} features substantially improves estimation robustness.  This matters because transmission time of intermediate activations is often a dominant term in E2E latency, and incorrect throughput assumptions can push the split decision toward suboptimal points.


Most prior SC evaluations focus on CNN backbones (e.g., VGG-style networks). In contrast, modern vision systems increasingly adopt transformer-based designs. Swin Transformer’s multi-stage hierarchy and progressive feature abstraction provide new partitioning opportunities, but also introduce new compute/activation size profiles that differ from CNNs. This paper extends adaptive splitting to Swin Transformer and demonstrates end-to-end feasibility for real-time video object detection in AI-on-RAN settings. 

Unlike many collaborative inference~\cite{DeViT_TMC_2024, boomerang} that rely on training multiple model variants (e.g., specialist/generalist ensembles), and often require sending richer information (privacy-related) to the network, our approach does not retrain or modify the base model. We execute an unmodified transformer and simply partition its forward pass across UE and Edge. 
This design preserves the key privacy advantage of split computing—\textit{raw sensor data never leaves the device}—while enabling adaptive offloading under time-varying 5G conditions. We also address a transformer-specific challenge: intermediate activations can be orders of magnitude larger than the input, making uplink transmission a bottleneck.
To tackle this, we (i) identify a set of \textit{deployment-friendly split candidates }aligned with Swin Transformer’s hierarchical stage/block boundaries, and (ii) introduce a lightweight \textit{activation compression pipeline} that rapidly compresses transmitted features without degrading end-to-end accuracy, thereby making transformer splitting viable for real-time operation.




Our main contributions are:
\begin {itemize}

    \item \textit{Feasibility of splitting Swin Transformer}: we extend throughput-aware adaptive splitting from CNNs to a Swin Transformer backbone and demonstrate practical split execution for transformer-based vision inference.
    
    \item \textit{Intermediate activation compression}: we propose an efficient compression pipeline for intermediate representation that substantially reduce uplink payload while maintaining model accuracy. 

    \item \textit{E2E real-time validation}: we implement and validate the full pipeline (adaptive splitting + transformer backbone) on a real-time detection workload, demonstrating robust performance under varying channel conditions.

    \item \textit{System-level latency reduction with distributed UPF (dUPF)}:
    We integrate dUPF into the edge AI inference path to reduce user-plane transport delay, further improving E2E latency for split inference.

    \item \textit{Comprehensive energy accounting (inference + 5G communication)}:
    We evaluate UE energy by jointly measuring on-device computation and 5G communication energy, providing a realistic assessment of the latency-energy-privacy trade-off.

    \item \textit{Testbed demonstration:} The proposed system is evaluated on an AI-RAN 5G testbed~\cite{Hariz_Infocom2026} based on the NVIDIA Aerial stack and was demonstrated at the AI-RAN Alliance booth at MWC Barcelona 2026~\footnote{\href{MWC Barcelona 2026 demo link}{https://youtu.be/yM4k4H9nPKM}}.

    
\end{itemize}

\section{Related Works}
\label{sec:relatedWork}

\subsection{ML model Splitting in Edge-Cloud Networks}

Split inference partitions DNNs between user devices and edge/cloud servers to trade off latency, energy, and privacy. 
Early work such as BranchyNet~\cite{BranchyNet_2016}, NeuroSurgeon~\cite{neurosurgeon}, JointDNN \cite{jointDnn}, as well as industrial systems such as Auto-Split~\cite{AutoSplit}, primarily focus on either DNN-based or CNN-based models with static or resource-aware partitioning strategies. 
Cooperative schemes like Boomerang~\cite{boomerang} further enable on-demand CNN inference by distributing computation across multiple edge resources. 
More recent works study adaptive partitioning under dynamic wireless conditions~\cite{binucci2024enablingedgeartificialintelligence, TamGlobecom2025}, jointly optimizing splitting points, communication, and energy, but largely assume CNN backbones and simplified channel models. Our prior work~\cite{TamGlobecom2025} introduced ML-based throughput-aware adaptive splitting over 5G, demonstrating gains for CNN-based vision tasks.

\subsection{Transformer and ViT-specific collaborative inference}
While most prior split inference work focuses on CNNs, transformer models pose new challenges due to attention complexity and different activation size profiles. 
Recent works investigate collaborative inference for Vision Transformers~\cite{ViT_2020}, including hierarchical decomposition and multi-tier execution across edge and near-edge accelerators. 
Frameworks such as DeViT~\cite{DeViT_TMC_2024} and FactionFormer~\cite{FactionFormer} exploit ViT structure to reduce computation or specialize models for edge scenarios, but typically rely on fixed or semi-static offloading strategies and do not integrate real-time wireless awareness

In contrast, this work (i) extends adaptive, throughput-aware split inference from CNNs to Swin Transformer backbone,
(ii) explicitly leverages the hierarchical Swin Transformer architecture for multiple feasible split boundaries, and
(iii) validating the approach in a real AI-RAN system, rather than simulation-only or edge–cloud abstractions.


\section{System Architecture}
\subsection{Architecture overview}

\begin{figure*}[!t]
    \centering
    \includegraphics[width=0.85\textwidth]{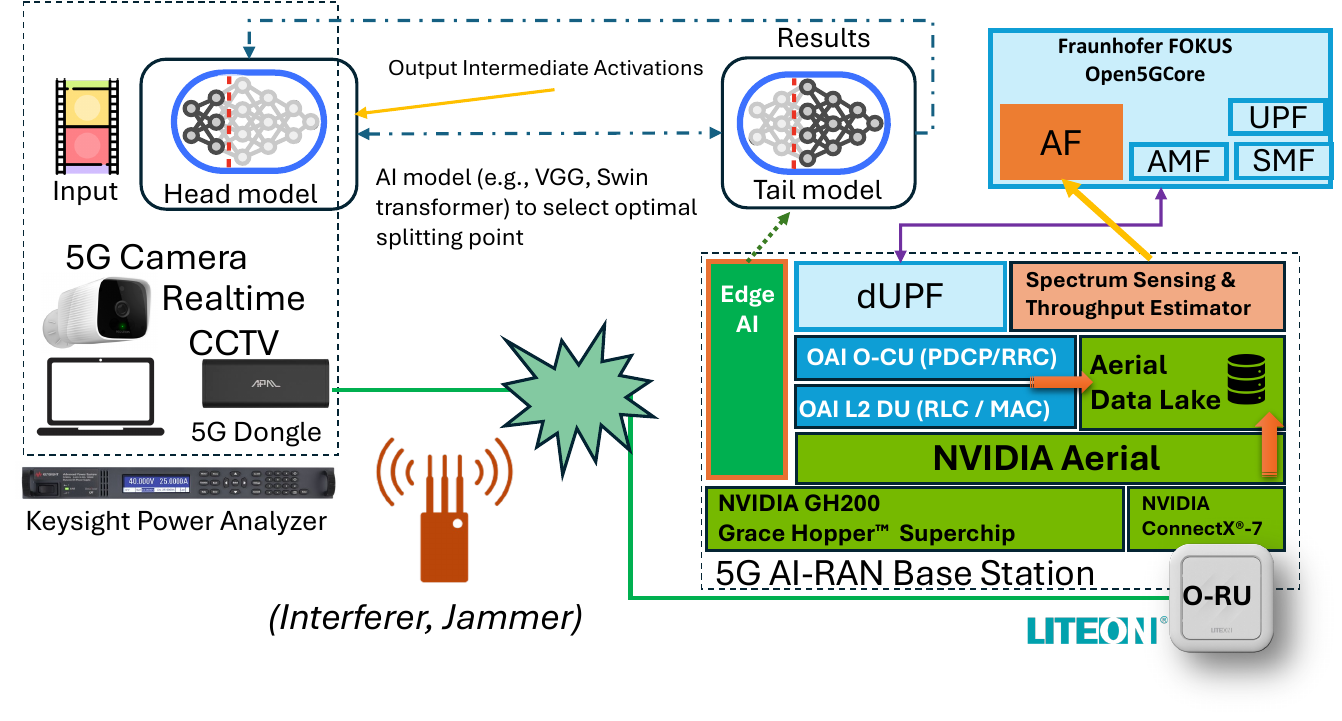}
    \vspace{-5mm}
    \caption{System Architecture of Adaptive Transformer Model Partitioning over AI-RAN Networks.}
    \vspace{-3mm}
    \label{fig:SystemArchitecture}
\end{figure*}

Fig.~\ref{fig:SystemArchitecture} illustrates the updated architecture of \textit{Adaptive Transformer Model Partitioning over AI-RAN networks}, integrating real-time video object detection, the NVIDIA Aerial 5G stack, and a dUPF enabled by Fraunhofer FOKUS Open5GCore.

The end-to-end vision pipeline follows a split-inference paradigm. A deep neural network (e.g., a Swin Transformer backbone with a detection head) is partitioned at layer $l$ into two components:
\begin{itemize}
    \item \textbf{Head model (layers $1$ to $l$)} runs on the UE.
    \item \textbf{Tail model (layers $l+1$ to $L$)} is executed on the Edge AI server.
\end{itemize}

During inference, the head model outputs \textit{intermediate data} that are transmitted over the uplink to the Edge. The wireless environment is time-varying---and may be affected by external interference or jamming---the system dynamically selects an optimal splitting point $l^*$ to maintain real-time performance while balancing E2E delay, privacy, and energy efficiency.

On the wireless network side, we deploy a 5G AI-RAN base station powered by the NVIDIA Aerial stack. The RAN includes the OAI O-CU (PDCP/RRC) and OAI L2 DU (RLC/MAC) components, which expose radio-layer observations leveraged for adaptive decision-making. The entire stack runs on an NVIDIA Grace Hopper GH200 platform with an NVIDIA ConnectX-7 NIC for hardware-accelerated processing, as shown in Fig.~\ref{fig:SystemArchitecture}. We use a commercial-grade LiteON O-RU, connected to the base station via the Open Fronthaul (OFH) interface.
5G Core is deployed in a separate server; implementation details are described in Section~\ref{subsec:dUPF}.
In parallel, the Aerial Data Lake continuously aggregates RAN and RF-related data (e.g., IQ and channel measurements) for monitoring, profiling, and model-driven optimization.

To support network awareness in real time, we incorporate a \textit{Spectrum Sensing \& Throughput Estimator} at the edge AI-RAN node. This module characterizes instantaneous channel quality and throughput variations, and its output is used to guide adaptive partitioning decisions, especially under interference-heavy conditions.

In addition, to quantify the energy impact on the UE side, we instrument the device using a Keysight Power Analyzer, enabling precise measurement of on-device computation energy and the extra energy cost induced by 5G transmission.

\subsection{5GCore and dUPF-enabled low-latency edge path}
\label{subsec:dUPF}

The 5G core network is implemented using Fraunhofer FOKUS Open5GCore\footnote{\href{Open5GCore}{https://open5gcore.org/}}, a standards-compliant 3GPP Releases 17/18 5G Core platform that provides essential core network functions, including the Access and Mobility Management Function (AMF), Session Management Function (SMF), and User Plane Function (UPF), etc. 
The platform supports flexible, and testbed-oriented  deployment, enabling the instantiation of dUPF instances at the network edge, e.g., collocated at AI-RAN node, to support local breakout N6 interface.

With dUPF deployment, Edge inference-related user-plane traffic is locally anchored and routed to the Edge AI service hosted on the same AI-RAN cluster, enabling low-latency edge computing. 
In contrast, Cloud services traffic continues to be routed through the centralized UPF in the core network. 
By steering latency-critical flows—such as intermediate activations and inference outputs—to the edge dUPF, while keeping normal user traffic (e.g., web browsing and cloud-bound services) on the central UPF, the system shortens the end-to-end path, reduces backhaul utilization, and improves responsiveness for real-time edge AI workloads.




\subsection{AF (Application Function)}

Following our prior work~\cite{TamGlobecom2025}, the Application Function (AF) employs a multi-objective optimization framework to dynamically select the model split point by jointly balancing E2E delay, privacy leakage, and UE energy under predefined constraints, with decisions adapted in real time based on RAN-provided throughput estimates.

\section{Splitting Swin Transformer}
\subsection{Swin Transformer}

\begin{figure*}[!t]
    \centering
    \includegraphics[width=\textwidth]{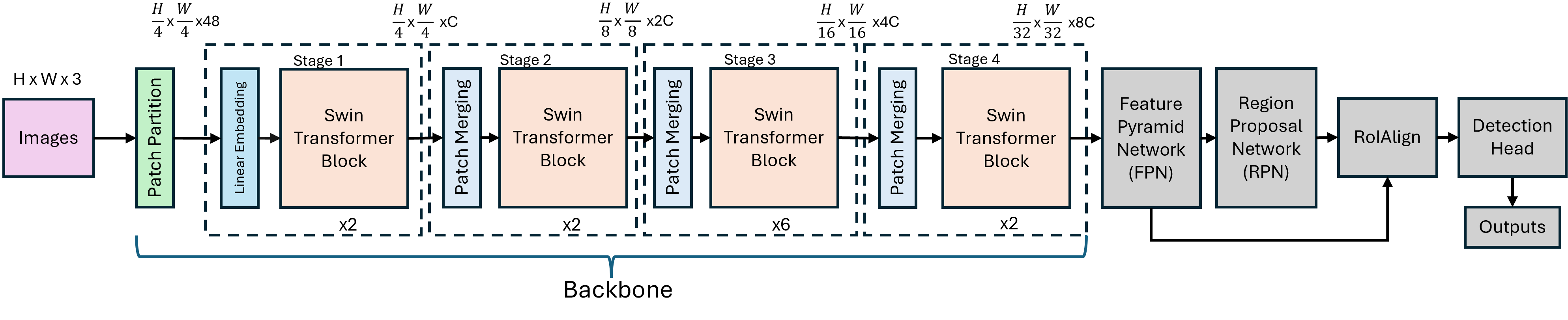}
    \caption{Object Detection Model with Swin Transformer Backbone.}
    \label{fig:mask_rcnn}
\end{figure*}

Fig.~\ref{fig:mask_rcnn} show the object detection architecture based on Swin Transformer backbone. Swin Transformer~\cite{liu2021swin} is a hierarchical vision transformer that employs shifted-window self-attention to achieve efficient, near-linear computational complexity when producing multi-scale feature representations for dense prediction tasks. 
Its stage-wise design---comprising patch embedding $\rightarrow$ repeated Swin blocks $\rightarrow$ patch merging between stages---naturally exposes partition boundaries where resolution and channel dimensions change, impacting both compute and activation size.

In this work, model partitioning is applied \textit{exclusively within the backbone}, as shown in Fig.~\ref{fig:mask_rcnn}. 
All downstream components, including the Feature Pyramid Network (FPN), Region Proposal Network (RPN), RoIAlign, and detection heads, are executed on the server whenever split inference is enabled. 
FPN aggregates multi-scale features produced by the backbone into a representation that improve detection across object sizes; 
RPN proposes candidate object regions;
and RoIAlign (Region of Interest) extracts fixed-size object-level features that are processed by lightweight detection heads for bounding box classification and localization. See Mask-RCNN paper~\cite{He_2017_ICCV} for details of this detection pipeline. 

By restricting partitioning to the feature extraction stage, this design maintains a stable detection pipeline on the server while enabling flexible split inference in dynamic 5G environments.


\subsection{Splitting points}

We define a set of deployment-friendly candidate split points $\mathcal{L}$ with predictable tensor shapes, including splits after patch embedding (early split with large activations), within Swin stages (fine-grained control), and after patch merging between stages (coarser resolution with smaller activations and stronger privacy).
In the current implementation, we restrict the partitioning strategy to Swin stage-level splitting points. 
Extending the framework to support finer-grained splitting points, such as intra-stage partitioning, is left for future work.

Despite these constraints on split-point selection, the size of intermediate activations remains substantially larger than the original input image. For example, while the input image occupies only 1.312~MB, intermediate features produced after a Swin block can reach 34--45~MB, exceeding the input size by more than 35$\times$, refer to Fig~\ref{fig:compression_reduction_pct}. 
This large expansion of feature size makes uplink transmission a dominant bottleneck in split inference and motivates the need for intermediate feature compression.
\subsection{Compression Intermediate Data}
\begin{figure}
    \centering
    \includegraphics[width=1\linewidth]{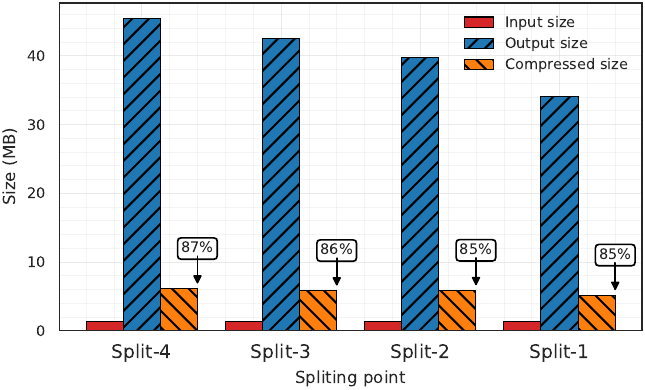}
    \caption{Intermediate data size (MB) vs compressed size (MB) across different splitting points.}
    \label{fig:compression_reduction_pct}
\end{figure}

As shown in Fig.~\ref{fig:compression_reduction_pct}, transmitting raw intermediate representations incurs significant wireless overhead.
To reduce this wireless transmission overhead, we introduce a compression technique to reduce the size of intermediate features including two steps: (1) quantizing the activations from FP32 to INT8, and (2) compressing the quantized features on the UE before transmission, followed by decompression on the server after reception using \texttt{zlib}. With this design, Fig.~\ref{fig:compression_reduction_pct} shows that the transmitted payload is reduced to about $5$--$6$~MB across splitting points, corresponding to a $\sim 85\%$--$87\%$ size reduction, which significantly shortens transmission time and improves end-to-end latency.

\section{Experimental Setup and Results}
\subsection{Experimental Setup}

For real-time video object detection, we adopt a Swin Transformer backbone as proposed in \cite{liu2021swin}, using the official implementation~\footnote{\href{Swin-Transformer-Object-Detection}{https://github.com/SwinTransformer/Swin-Transformer-Object-Detection}}.
The detection pipeline, refer Fig.~\ref{fig:mask_rcnn}, follows a standard Transformer-based architecture, consisting of a Swin Transformer backbone for feature extraction and a detection head (e.g., Feature Pyramid Network (FPN) + detection head) for object localization and classification. To enable optimized deployment over a 5G network, we partition the backbone into two components and execute them across the UE and the edge server:
and split the model into two components to enable optimized deployment over a 5G network:
\begin{itemize}
    \item \textbf{5G UE}: A wireless 5G endpoint (e.g., a 5G camera) that hosts the \textbf{head model}. In our experiments, we emulate the UE using a GPU-free laptop equipped with 13\textsuperscript{th} Gen Intel\textsuperscript{\textregistered} Core\textsuperscript{\texttrademark} i9-13900H processor and 32\,GB RAM,
        connected to a Pegatron 5G dongle in pass-through mode for 5G network access.
    \item \textbf{Edge Server (AI-RAN node)}: A high-performance NVIDIA GH200 platform that allocates GPU resources via MIG-based partitioning. Following the AI-RAN node design in~\cite{Hariz_Infocom2026}, this node co-locates \textbf{tail model} execution with 5G RAN base station workload, enabling joint RAN processing and Edge AI inference on the same platform.
\end{itemize}

Swin Transformer offers 4 potential splitting points, corresponding to outputs of its four stages separated by patch-merging operations.
These stage-level boundaries offer predictable tensor shapes and well-defined semantic abstraction levels, making them suitable for deployment-oriented split inference. 
By splitting at these points, we can systematically analyze the trade-offs among intermediate feature size, computational load, and privacy leakage—quantified via the distance correlation between the transmitted representation and the original input—under dynamic 5G network conditions.


To compare \textit{Edge AI over dUPF} and \textit{Cloud AI over centralized UPF (cUPF)}, Cloud AI is deployed on a separate server with inference traffic routed through the cUPF. The backhaul between the cUPF and Cloud AI is emulated using Linux Traffic Control (\texttt{tc}) by injecting a one‑way delay of $100$~ms with $5$~ms jitter on each direction.


For repeatable and fair benchmarking, we use a fixed 20-second prerecorded video clip to profile privacy leakage \(P(l)\) and UE energy consumption \(E_{\text{UE}}(l)\) at each split point. Experiments are conducted under both normal and controlled interference scenarios, where jamming power is increased from \(-40\,\mathrm{dB}\) to \(-5\,\mathrm{dB}\), and UE-side latency, transmission, and energy metrics are logged for analysis.

\subsection{Experimental Results}

\begin{figure}
    \centering
    \includegraphics[width=0.95\linewidth]{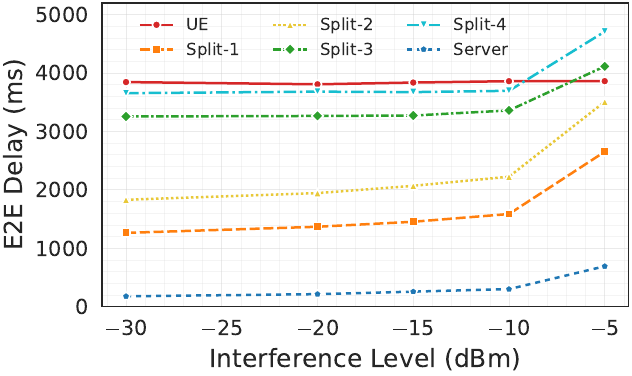}
    \caption{E2E Delay (ms) over different splitting points (i.e., UE, Split 1--4, Server) under different interference levels.}
    \label{fig:e2edelay_vs_interference}
\end{figure}

\subsubsection{Analysis of E2E Delay Under Varying Scenarios}
As shown in Fig.~\ref{fig:e2edelay_vs_interference}, fully offloading raw input image to the Edge Server (blue dashed line) consistently achieves the lowest E2E delay, with a mean E2E delay of $327.6$~ms, compared to $3842.7$~ms for fully local UE inference---corresponding to an $11.7\times$ reduction in E2E delay.
Under low-to-moderate interference levels ($-30$~dB to $-10$~dB), all split configurations outperform UE-only execution; for example, Split-1 achieves $1262.9$~ms at $-30$~dB and remains $1586.1$~ms at $-10$~dB. In contrast, UE-only E2E delay stays nearly constant (from $3845.6$~ms at $-30$~dB to $3860.8$~ms at $-10$~dB), indicating that computation dominates local execution. However, when interference becomes severe ($-5$~dB), split strategies experience a sharp E2E delay increase due to degraded uplink transmission of intermediate activations: Split-1 rises to $2652.8$~ms and deeper splits can exceed UE-only inference (e.g., Split-3 at $4114.6$~ms and Split-4 at $4710.0$~ms vs.\ $3862.0$~ms on the UE). Even server-only inference is affected at $-5$~dB, increasing to $691.1$~ms, but it still remains significantly faster than all split and local baselines.

\begin{figure}[ht]
    \centering
    \includegraphics[width=1\linewidth]{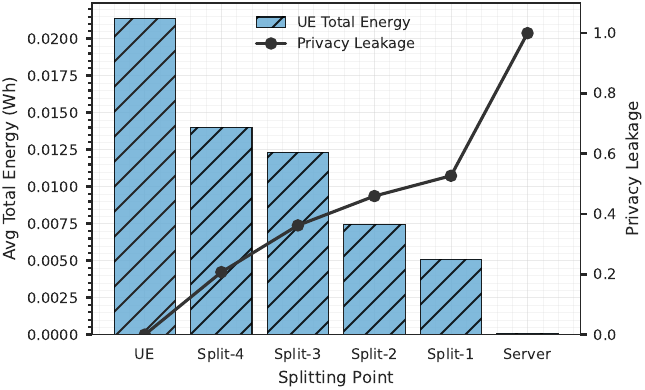}
    \caption{UE's total energy consumption, including Communication \& Compute, (in bar graph) and UE's Privacy leakage (in line graph) across different splitting points.}
    \label{fig:privacy_energy_tradeoff}
\end{figure}

\subsubsection{UE Energy Breakdown: 5G Transmission vs. Inference}

\begin{figure}[ht]
    \centering
    \includegraphics[width=1\linewidth]{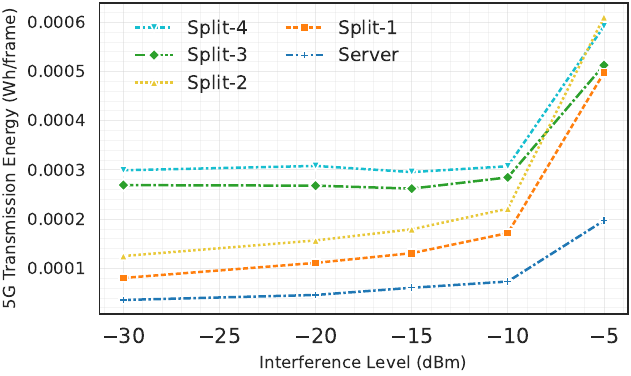}
    \caption{UE's Energy for 5G Transmission over different splitting points under different interference levels.}
    \label{fig:network_energy_vs_interference}
\end{figure}

Fig.~\ref{fig:network_energy_vs_interference} shows the UE energy consumed for 5G transmission under varying interference levels. As interference increases, the 5G dongle draws more power to transmit the same intermediate features, leading to higher transmission energy per frame. This increase is moderate at low to medium interference but becomes pronounced under severe interference (e.g., $-5,\mathrm{dB}$), reflecting the additional radio effort required to sustain reliable uplink transmission.

\begin{figure}[ht]
    \centering
    \includegraphics[width=1\linewidth]{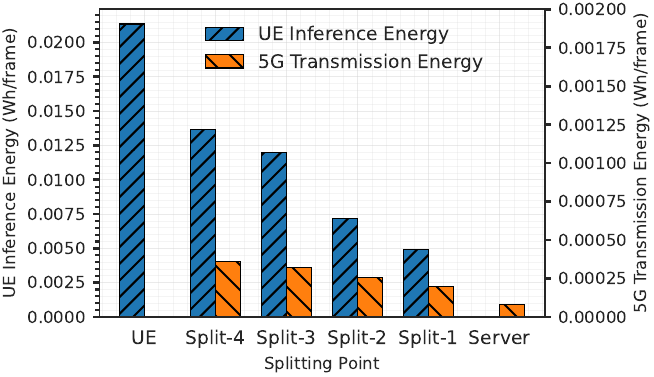}
    \caption{UE's Energy consumed for Inference (left axis) and for 5G Transmission (right axis) over different splitting points.}
    \label{fig:energy_contribution_breakdown}
\end{figure}


Fig.~\ref{fig:energy_contribution_breakdown} shows that UE inference energy dominates UE 5G transmission energy, where the latter is averaged over interference levels. Total UE energy decreases from $0.0213$~Wh/frame for fully local UE inference to $0.0051$~Wh/frame for a shallow split (Split‑1, $76.1\%$ reduction) and to $0.0001$~Wh/frame for Server‑only execution ($99.5\%$ reduction). 
In contrast, 5G transmission energy remains $25\times$–$50\times$ smaller than inference energy, \textit{indicating that computation offloading is the primary driver of UE energy savings}.



\subsubsection{Privacy leakage}
As depicted Fig.~\ref{fig:privacy_energy_tradeoff}, Privacy leakage increases with deeper offloading, rising from $0$ under local processing to $0.527$ for Split-1 and peaking at $1.0$ for server-only inference, where raw inputs are transmitted to the server. This follows our distance-correlation metric: earlier-layer outputs preserve more input structure and thus remain more correlated with the original content, while deeper features are more abstract and leak less. Unlike E2E delay and energy, privacy leakage is not affected by wireless conditions because it depends on \emph{what} is transmitted (raw input vs.\ activations at splitting point $l$), not on the channel state.

\subsubsection{Performance under dUPF vs cUPF}
\label{subsubsec:cUPF_dUPF}

\begin{figure}[htb]
    \centering
    \includegraphics[width=1\linewidth]{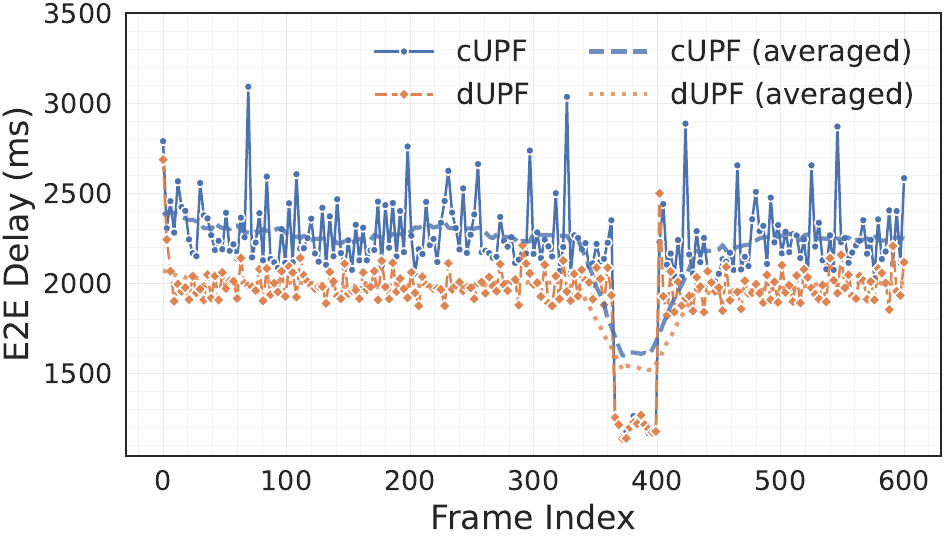}
    \caption{E2E delay comparison between Cloud AI over cUPF and Edge AI over dUPF.}
    \label{fig:e2e_delay_cupf_dupf}
\end{figure}

Fig.~\ref{fig:e2e_delay_cupf_dupf} compares the E2E delay between Cloud AI over cUPF (blue line) and Edge AI over dUPF (orange line), with dashed lines show moving average of raw values. Overall, cUPF exhibits consistently higher delay, with a mean E2E delay of $2199.73$~ms compared to $1944.13$~ms for dUPF---an increase of $255.59$~ms (about $13.1\%$). 
In addition to the lower mean delay, dUPF shows smaller delay fluctuations (std $211.77$~ms vs.\ $310.58$~ms for cUPF), indicating reduced jitter and more \textit{stable runtime behavior}. 
This is because, under cUPF, inference traffic must traverse the external Internet and core-network backbone, incurring longer propagation delays and unpredictable jitter, whereas dUPF locally anchors traffic at the edge, resulting in lower latency and more stable performance. 
These results suggest that placing the UPF closer to the edge (dUPF) improves both E2E delay and stability for real-time video inference over 5G.

\subsection{Summary and Discussion}
Our measurements show that split inference can significantly reduce E2E delay under low-to-moderate interference, but may degrade under extreme interference when uplink transmission of intermediate activations dominates; meanwhile, offloading reduces UE energy but increases privacy leakage, revealing a clear E2 delay--privacy--energy trade-off. We also find that deploying dUPF closer to the edge improves both mean E2E delay and stability compared to a cUPF.

Rather than seeking a globally optimal split, this work studies the feasibility of representative split points to expose key trade‑offs, motivating future automated split selection that accounts for network uncertainty, activation compression, and edge resource orchestration to enable robust online mode switching.



\section{Conclusions}
\label{sec:conclusions}
In this paper, we studied split inference of a Swin Transformer for real‑time video object detection over dynamic 5G networks, deployed between a 5G UE and an edge server co‑located with a 5G AI‑RAN base station. Our experimental results provide practical insights into deploying transformer‑based split inference under time‑varying wireless conditions and motivate future work on automated split‑point selection, robust online execution‑mode switching, and privacy‑preserving, no‑retraining, accuracy‑preserving activation compression for transformer backbones.


\textbf{Acknowledgment}: 
This research is supported 
by the National Research Foundation (NRF), Singapore and Infocomm Media Development Authority (IMDA) under its Future Communications Research \& Development Programme.
Any opinions, findings, conclusions, or recommendations expressed in this material are those of the authors and do not reflect the views of NRF and IMDA Singapore. 

\bibliographystyle{IEEEtran}
\bibliography{ref}

\end{document}